\documentstyle[12pt]{article}
\begin{document}
\def\beq{\begin{equation}}
\def\eeq{\end{equation}}
\def\bea{\begin{eqnarray}}
\def\eea{\end{eqnarray}}
\def\ve{\vert}
\def\vel{\left|}
\def\ver{\right|}
\def\nnb{\nonumber}
\def\ga{\left(}
\def\dr{\right)}
\def\aga{\left\{}
\def\adr{\right\}}
\def\rar{\rightarrow}
\def\nnb{\nonumber}
\def\la{\langle}
\def\ra{\rangle}
\def\ba{\begin{array}}
\def\ea{\end{array}}
\def\tep{$B \rar K^* \ell^+ \ell^-$}
\def\tepm{$B \rar K^* \mu^+ \mu^-$}
\def\tept{$B \rar K^* \tau^+ \tau^-$}
\def\ds{\displaystyle}

\title{ {\small {\bf RARE \tep DECAY, TWO HIGGS DOUBLET MODEL, AND LIGHT
CONE $QCD$ SUM RULES} } }

\author{\vspace{1cm}\\
{\small T. M. AL\.{I}EV \thanks
{e-mail: taliev@rorqual.cc.metu.edu.tr}\,\,,
M. SAVCI \thanks
{e-mail: savci@rorqual.cc.metu.edu.tr}\,\,,
A. \"{O}ZP\.{I}NEC\.{I} \thanks
{e-mail: e100690@orca.cc.metu.edu.tr}} \\
{\small Physics Department, Middle East Technical University} \\
{\small 06531 Ankara, Turkey} \\
\vspace{5mm}\\  
{\small H. KORU}\\
{\small Physics Department, Gazi University} \\
{\small 06460 Ankara, Turkey} } 

\date{}

\begin{titlepage}
\maketitle
\thispagestyle{empty}

\begin{abstract}
\baselineskip  0.7cm

The decay width, forward-backward asymmetry and lepton
longitudinal and transversal polarization for the exclusive \tep decay in a two Higgs
doublet model are computed. It is shown that all these quantities are 
 very effective tools for establishing new physics.
\end{abstract}

\vspace{1cm}
\end{titlepage}

\section{Introduction} 

Experimental discovery of the inclusive and exclusive $B \rar X_s \gamma$
and $B \rar K^* \gamma$ \cite{R1} decays stimulated the study of rare $B$
meson decays in a new manner. These decays take place via
flavor-changing neutral current (FCNC) $b \rar s$ transitions which are absent in
the Standard Model (SM) at tree level and appear only at the loop level.
Therefore the study of
these decays can provide sensitive tests for investigating the structure of the
SM at the loop level, searching for new physics
beyond the SM \cite{R2}.

Currently, the main interest is focused on the rare $B$ meson decays 
for which the SM predicts large branching ratios and that can be potentially
measurable in the near future. The rare \tep ($\ell=e,~\mu,~\tau$) decays are
such decays. For these decays the experimental situation is very promising
\cite{R3} with $e^+ e^-$ and hadron colliders focusing only on
the observation of exclusive modes with $l = e,~\mu$ and $\tau$ as
the final states. At the quark level, the process \tep \, is
described by the  $b \rar s l^+ l^-$ transition. 
In literature [4-17] this transition has been investigated extensively 
in both SM and two Higgs doublet model  (2HDM). 
It is well known that in the 2HDM, the up type quarks acquire their masses
from
Yukawa couplings to the Higgs doublet $H_2$ (with the vacuum expectation
value $v_2$) and down type quarks and
leptons acquire their masses from Yukawa couplings to the other Higgs
doublet
$H_1$ (with the expectation value $v_1$).  In 2HDM there exist five physical
Higgs fields: neutral scalar $H^0$, $h^0$, neutral pseudoscalar $A^0$ and
charged Higgs bosons $H^\pm$. Such a model occurs as a natural feature of
the supersymmetric models \cite{R18}. In these models the interaction
vertex of the Higgs boson and
fermions depends on the ratio $tan \,\beta = \frac{\ds{v_2}}{\ds{v_1}}$
which is
a free parameter in the model. The constraints on $tan \,\beta$ are usually
obtained from $B-\bar B,~ K-\bar K$ mixing, $b \rar s \gamma$ decay width,
semileptonic decay $b \rar c \tau \bar \nu_\tau$ and is given by
\cite{R19,R20}:

\beq
0.7 \le tan \, \beta \le 0.6 \left( \frac{m_{H^+}}{1~GeV} \right)
\eeq
(the lower bound $m_{H^+} \ge 200~GeV$ is obtained in \cite{R20}).

In 2HDM the charged and neutral Higgs interactions with fermions induce new
contributions to the FCNC processes. For $b \rar s \ell^+ \ell^-$ ($\ell =
e,~\mu$) decay the
contributions from neutral Higgs boson exchange are 
usually neglected, due to the fact that the interaction of the neutral Higgs and
leptons  is proportional to the lepton mass. But for the $b \rar s \tau^+ \tau^-$
decay the mass of the $\tau$ lepton is not too small compared to the mass of
the $b$-quark,
therefore the neutral Higgs boson exchange diagrams can give
considerable contributions to the process. Recently it has been emphasized by
Hewett \cite{R21} that $\tau$ lepton longitudinal polarization is an important
observable and may be accessible in the $B \rar X_s \tau^+ \tau^-$ decay.
In \cite{R22} it is shown that in addition to longitudinal component  $P_L$ of
polarization of the $\tau$ lepton two other orthogonal components 
$P_T$, which lies in the decay plane, and $P_N$ which is perpendicular to the 
decay plane, are also very 
significant for the $\tau^+~\tau^-$ channel, since they are proportional 
to the mass of the $\tau$ lepton. The polarization components
$P_L,~P_T$ and $P_N$ involve different combinations of the Wilson
coefficients $C_7,~C_9^{eff}$ and $C_{10}$ (see below) and hence
contain independent information. Therefore, polarization effects can play
a central role for the investigation of the structure of the SM and for
establishing new physics beyond it.

In calculating the branching ratios and other observables at hadronic level,
e.g., for \tep decay, we face  the problem of computing the matrix
element of the effective Hamiltonian responsible for this decay, between $B$
and $K^*$ states. This problem is related to the non-perturbative sector
of QCD and it can be solved only by means of a non-perturbative approach.

These matrix elements have been investigated in the framework of different
approaches such as chiral theory \cite{R23}, three point QCD sum rules 
method \cite{R24}, relativistic quark model by the light-front formalism 
\cite{R25}, effective heavy quark theory \cite{R26} and light cone QCD 
sum rules \cite{R27}. The aim of the present work is to calculate these
matrix elements using the light cone QCD sum rules formalism in the framework 
of the 2HDM, taking into account the newly appearing operators due to the
neutral Higgs exchange diagrams, 
and to study the forward-backward asymmetry and final lepton polarization for the 
exclusive \tep decay. Taking into account the additional neutral Higgs boson
exchange diagrams, the effective Hamiltonian is calculated in \cite{R28} as
\bea
{\cal H}_{eff} = \frac{4 G_F}{\sqrt 2} V_{tb} V^*_{ts} \left\{ \sum_{i=1}^{10}  
C_i( \mu ) O_i( \mu ) + \sum_{i=1}^{10} C_{Q_i}( \mu ) Q_i( \mu ) \right\} ~,
\eea
where the first set of operators in the curly brackets describe the effective Hamiltonian
responsible for the $b \rar s l^+ l^-$ decay in the SM. The value of the
corresponding Wilson coefficients includes the contribution of the diagrams
with $H^\pm$ running in the loop (see \cite{R7}), i.e.
\bea
C_7 (M_W) &=& C_7^{SM} (M_W)+ C_7 ^{H^-} (M_W)~, \nnb  \\
C_9 (M_W) &=& C_9^{SM} (M_W)+ C_9 ^{H^-} (M_W)~, \nnb   \\
C_{10} (M_W) &=& C_{10}^{SM} (M_W)+ C_{10}^{H^-} (M_W)~, \nnb
\eea
where 
\bea
C_7^{H^-} (M_W)&=& -\frac{y}{2} \left[ \frac{5 y - 3}{6 (y-1)^2} - \frac{3 y -2}
{3 (y-1)^3} ln \, y \right] - \nnb \\
&-& \frac{1}{6} ctg^2 \beta \, y \left[ \frac{8 y^3 + 5 y - 12}{(y-1)^3} -
\frac{3 y^2 - 2 y}{2 (y-1)^4} ln \, y \right]~,\nnb \\
C_9^{H^-} (M_W)&=& \frac{1 - 4 sin^2 \theta_W}{sin^2 \theta_W} \left[ ctg^2 \beta
\frac{x y }{8} \left( \frac{1}{y-1} - \frac{1}{(y-1)^2} ln \, y \right)
\right] - \nnb \\
&-& ctg^2 \beta \, y \left[ \frac{47 y^2 - 79 y + 108}{108 (y-1)^3} -
\frac{3 y^3 - 6y + 4}{18 (y-1)^4} ln \, y \right]~, \nnb \\
C_{10}^{H^-} (M_W) &=&  \frac{1}{sin^2 \theta_W} ctg^2 \beta \frac{x y}{8}
\left[ - \frac{1}{y-1} + \frac{1}{(y-1)^2} ln \, y \right]~. \nnb 
\eea
The explicit forms of $C_7^{SM} (M_W)$, $C_9^{SM} (M_W)$, and $C_{10}^{SM}
(M_W)$ and the corresponding operators can be found in \cite{R7}.

The second set of operators in the brackets, whose explicit forms are presented
in \cite{R28}, come from the exchange of the neutral Higgs bosons. 
The corresponding Wilson coefficients are:
\newpage
\bea
C_{Q_1} ( m_W ) &=& \frac{m_b m_\ell}{m_{h^0}^2} tan^2 \, \beta
\frac{1}{sin^2 \theta_W} \frac{x}{4} \Bigg\{ \left(sin^2 \alpha + h\, cos^2
\alpha \right) f_1 (x,y) + \nnb \\
&+& \left[ \frac{m_{h^0}^2}{m_W^2} + \left(sin^2 \alpha + h\,
cos^2\alpha \right)(1-z) \right] f_2(x,y) +  \nnb \\
&+& \frac{sin^2 2 \alpha}{2 m_{H^\pm}^2} \left[m_{h^0}^2 -
\frac{(m_{h^0}^2 + m_{H^0}^2)^2}{2 m_{H^0}^2} \right] f_3 (y) \Bigg\}~,\\
C_{Q_2} (m_W) &=& \frac{m_b m_\ell}{m_{H^\pm}^2} tan^2 \, \beta \left\{ f_1(x,y) +
\left[1+ \frac{m_{H^\pm}^2 - m_{A^0}^2}{m_W^2} \right] f_2(x,y) \right\}~,\\
C_{Q_3} (m_W) &=& \frac{m_b e^2}{m_\ell g^2} \Bigg[ C_{Q_1} (m_W) + C_{Q_2}
(m_W) \Bigg] ~,\\
C_{Q_4} (m_W) &=& \frac{m_b e^2}{m_\ell g^2} \Bigg[ C_{Q_1} (m_W) -
C_{Q_2}(m_W) \Bigg]~, \\ 
C_{Q_i}(m_W) &=& 0 ~~~~~~ i =5, \ldots,10~,
\eea
where
$$x = \frac{m_t^2}{m_W^2}~,~~~~y=\frac{m_t^2}{m_{H^\pm}^2}~,~~~~z=
\frac{x}{y}~,~~~~h=\frac{m_{h^0}^2}{m_{H^0}^2}~, $$
$$f_1 (x,y) = \frac{x\, lnx}{x-1} - \frac{y\, lny}{y-1}~,~~~~ f_2(x,y) =
\frac{x\, lny}{(z-x)(x-1)} + \frac{lnz}{(z-1)(x-1)}~,$$
$$f_3(y) = \frac{1 -y + y\, lny}{(y-1)^2}~.$$
The QCD
correction to the Wilson coefficients $C_i(m_W)$ and $C_{Q_i}(m_W)$ can be
calculated using the renormalization group equations.
In \cite{R28} it was shown that the operators $O_9$ and $O_{10}$ do not
mix with $Q_i~(i=1, \ldots,10)$, so that the Wilson coefficients $C_9$ and
$C_{10}$ remain unchanged and their values are the same as in the SM. Their explicit
forms can be found in \cite{R28}, where it is also shown
that $O_7$ can mix with $Q_i$.
But additional terms due to this mixing
can  safely be neglected since the corrections to the
SM value of $C_7$ arising from these terms are less than $5 \%$ when $tan \,
\beta \le 50$.

Moreover the operators $O_i~(i=1, \ldots,10)$ and $Q_i~(i=3, \ldots,10)$ do
not mix with $Q_1$ and $Q_2$ and also there is no mixing between $Q_1$ and
$Q_2$. For this reason the evolutions of the coefficients $C_{Q_1}$ and
$C_{Q_2}$ are controlled by the anomalous dimensions of $Q_1$ and $Q_2$
respectively:
$$C_{Q_i} (m_b) = \eta^{-\gamma_Q / \beta_0} C_{Q_i} (m_W)~,~~~i=1,~2 ,$$
where $\gamma_Q = -4 $ is the anomalous dimension of the operator $\bar s_L b_R$
\cite{R29}.

Neglecting the strange quark mass, the matrix element for $b \rar s \ell^+
\ell^-$ decay is \cite{R28}:
\bea
{\cal M} &=& \frac{G_F \alpha}{2\sqrt 2 \pi} V_{tb} V^*_{ts} \Bigg\{ C_9^{eff}
\bar s \gamma_\mu (1- \gamma_5) b \, \bar \ell \gamma^\mu \ell + C_{10} \bar s
\gamma_\mu (1- \gamma_5) b \, \bar \ell \gamma^\mu \gamma_5 \ell -  \nnb \\
&-& 2C_7\frac{m_b}{q^2}\bar s i \sigma_{\mu \nu}q^\nu (1+\gamma_5)  b  \, 
\bar \ell \gamma^\mu \ell + C_{Q_1} \bar s (1 + \gamma_5) b \bar \ell \ell +
C_{Q_2} \bar s (1+\gamma_5) b  \bar \ell \gamma_5 \ell \Bigg\}~,
\eea
where $q^2$ is the invariant dileptonic mass. Here the coefficient
$C_9^{eff}(\mu,q^2) \equiv C_9( \mu) + Y ( \mu,~p^2)$, where the function $Y$
contains the contributions from the one loop matrix element of the
four-quark operators  \cite{R7,R30,R31}. It is clear that the last two terms
in eq.(8) can give considerable contribution only for the \tept decay mode, since
$C_{Q_1}$ and $C_{Q_2}$ are proportional to the lepton mass $m_\ell$.
In addition to the short distance contributions, it is possible to take 
into account the long distance effects 
associated with real $c\bar c$ in the intermediate states, i.e., the 
cascade process $B \rar K^* J/\psi(\psi^\prime)\rar K^* \ell^+ \ell^-$.
These contributions are taken into account by introducing a Breit-Wigner
form of the resonance propagator and this procedure leads to an additional
contribution to $C_9^{eff}$ of the form \cite{R32}
\bea
-\, \frac{3 \pi}{\alpha^2} \sum_{V=
  J/\psi,~ \psi^\prime,\ldots} \frac{m_V \Gamma(V \rar \ell^+
\ell^-)}{(q^2 - m_V^2) - i m_V \Gamma_V}~.
\nnb
\eea
From eq.(8) it follows that, in order to calculate the branching ratio for the
exclusive  \tept decay, the matrix elements
$ \la K^* \vel \bar s \gamma_\mu (1- \gamma_5) b \ver B \ra$,
$\la K^* \vel \bar s i \sigma_{\mu \nu} q^\nu (1+\gamma_5) b \ver B \ra$,
and $ \la K^* \vel \bar s (1+\gamma_5) b \ver B \ra$ have to be calculated.
These matrix elements can be written in terms of the form factors in the
following way:
\bea
\lefteqn{
\la K^* (p, \epsilon) \vel \bar s \gamma_{\mu}( 1- \gamma_5) b \ver B(p+q)
\ra = } \nnb \\
&-&  \epsilon_{\mu \nu \rho \sigma} \epsilon^{*\nu} p^\rho q^\sigma
\frac{2 V(q^2)}{m_B + m_{K^*}} - 
i \epsilon_\mu^* ( m_B + m_{K^*}) A_1(q^2)
+ i (2 p + q)_\mu (\epsilon^* q) \frac{A_2(q^2)}{m_B + m_{K^*}} + \nnb \\
&+& i q_\mu (\epsilon^* q) \frac{2 m_{K^*}}{q^2} \left[ A_3(q^2) - A_0 (q^2) \right]
~, \\ \nnb \\ \nnb \\
\lefteqn{
\la K^* (p, \epsilon) \vel \bar s i \sigma_{\mu \nu} q^\nu (1+ \gamma_5) b
\ver B(p+q) \ra = } \nnb \\
&& 4 \epsilon_{\mu \nu \rho \sigma} \epsilon^{* \nu} p^\rho
q^\sigma T_1 (q^2)
+  2 i \left[ \epsilon_\mu^* (m_B^2 - m_{K^*}^2) - (2 p + q)_\mu (\epsilon^* q)
\right] T_2 (q^2) + \nnb \\
&+&  2 i (\epsilon^* q) \left[ q_\mu - (2 p+q)_\mu \frac{q^2}{m_B^2 - m_{K^*}^2}
 \right] T_3 (q^2)~,
\eea
where $\epsilon$ is the polarization vector of $K^*$ meson. 

To
calculate the matrix element $\la K^* \ve \bar s (1+ \gamma_5) b \ve B
\ra$ we multiply both sides of eq.(9) by $q_\mu$ and use the equation of
motion. Neglecting the mass of the strange quark, we get:
\bea
\la K^*(p, \epsilon) \vel \bar s (1+ \gamma_5) b \ver B(p+q) \ra &=& 
\frac{1}{m_b} \Bigg\{ -i
(\epsilon^* q) ( m_B + m_{K^*}) A_1 (q^2) + \nnb \\
&+& i ( \epsilon^* q) (m_B - m_{K^*}) A_2(q^2) + \nnb \\
&+& i ( \epsilon^* q) 2 m_{K^*} \left[ A_3 (q^2) - A_0 (q^2) \right]
\Bigg\}~.
\eea
Using the equation of motion, the formfactor $A_3(q^2)$ can be written as a
linear combination of the formfactors $A_1$ and $A_2$ (see \cite{R24}):
\bea
A_3(q^2) = \frac{m_B + m_{K^*}}{2 m_{K^*}} A_1(q^2) - \frac{m_B - m_{K^*}}{2
m_{K^*}} A_2(q^2)~.
\eea
Making use of this relation we obtain:
\bea
\la K^*(p, \epsilon) \vel \bar s (1+ \gamma_5) b \ver B(p+q) \ra = \frac{2 m_{K^*}}{m_b}
\left\{-i (\epsilon^* q) A_0 (q^2) \right\}~.
\eea
From eqs.(8), (9), (10) and (13) we obtain for the matrix element of the  
\tep decay:
\bea
{\cal M} &=& \frac{G \alpha}{2 \sqrt 2 \pi} V_{tb} V_{ts}^*  \Bigg\{ \bar \ell
\gamma^\mu
\ell \left[ 2 A \epsilon_{\mu \nu \rho \sigma} \epsilon^{* \nu} p^\rho q^\sigma + i
B_1 \epsilon^*_\mu - i B_2 ( \epsilon^* q) ( 2 p + q)_\mu - i B_3 (\epsilon^* q)
q_\mu \right] + \nnb \\
&+& \bar \ell \gamma^\mu \gamma_5 \ell \left[ 2 C \epsilon_{\mu \nu \rho
\sigma}\epsilon^{* \nu} p^\rho q^\sigma + i D_1 \epsilon^*_\mu - i D_2 (
\epsilon^* q) ( 2 p + q)_\mu - i D_3 (\epsilon^* q) q_\mu \right] + \nnb \\
&+& i F_1 (\epsilon^* q ) \bar \ell \ell + i F_2 (\epsilon^* q) \bar \ell
\gamma_5 \ell \Bigg\}~,
\eea
where
\bea
A &=& C_9^{eff} \frac{V}{m_B + m_{K^*}} + 4 C_7 \frac{m_b}{q^2} T_1~, \nnb
\\ \
B_1 &=& C_9^{eff} (m_B + m_{K^*}) A_1 + 4 C_7 \frac{m_b}{q^2} (m_B^2 -
m_{K^*}^2) T_2~,  \nnb \\ \nnb \\ 
B_2 &=& C_9^{eff} \frac{A_2}{m_B + m_{K^*}} + 4 C_7 \frac{m_b}{q^2} \ga T_2 +
 \frac{q^2}{m_B^2 - m_{K^*}^2} T_3 \dr~,  \nnb \\ \nnb \\
B_3 &=& -C_9^{eff}\frac{ 2 m_{K^*}}{  q^2}(A_3 - A_0) + 4 C_7 
\frac{m_b}{q^2}T_3~,  \nnb \\ \nnb \\
C &=& C_{10} \frac{V}{m_B + m_{K^*}}~,  \nnb \\ \nnb \\  
D_1 &=& C_{10} (m_B + m_{K^*}) A_1~,  \nnb \\ \nnb \\    
D_2 &=& C_{10} \frac{A_2}{m_B + m_{K^*}}~,  \nnb \\ \nnb \\
D_3 &=& C_{10} \frac{2 m_{K^*}}{q^2} (A_3 - A_0)~, \nnb  \\
F_1 &=& C_{Q_1} \frac{2 m_{K^*}}{m_b} A_0~, \nnb \\
F_2 &=& C_{Q_2} \frac{2 m_{K^*}}{m_b} A_0~.
\eea
The formfactors $V,~A_1,~A_2,~A_0,~T_1,~T_2$ and $T_3$ are calculated
in the framework of light-cone QCD sum rules in \cite{R27} and their $q^2$ 
dependence, to a good accuracy, can be represented in the following pole
form:

\newpage

\bea
V(q^2) &=& \frac{0.55}{\displaystyle{\ga 1 - \frac{q^2}{30} \dr^2}}~,~~~
A_0(q^2) = \frac{0.26}{\displaystyle{\ga 1 - \frac{q^2}{45} \dr^2}}~, \nnb \\
A_1(q^2) &=& \frac{0.36}{\displaystyle{\ga 1 - \frac{q^2}{64} \dr^2}}~,~~~ 
A_2(q^2) = \frac{0.40}{\displaystyle{\ga 1 - \frac{q^2}{33} \dr^2}}~, \nnb \\ 
T_1(q^2) &=& \frac{0.18}{\displaystyle{\ga 1 - \frac{q^2}{31} \dr^2}}~,~~~ 
T_2(q^2) = \frac{0.18}{\displaystyle{\ga 1 - \frac{q^2}{64.4} \dr^2}}~, \nnb \\ 
T_3(q^2) &=& \frac{0.14}{\displaystyle{\ga 1 - \frac{q^2}{32.5} \dr^2}}~, 
\eea
which we will use in further numerical analysis (for $A_3$ we use eq.(12)).

Using eq.(14) and performing summation over final lepton polarization, 
we get for the double differential decay rate:
\bea
\frac{d \Gamma}{d q^2 dz} &=& \frac{G^2 \alpha^2
\vel V_{tb} V_{ts}^* \ver^2\lambda^{1/2} v}{2^{12}
\pi^5 m_B} \Bigg\{ 2 \lambda m_B^4 \Bigg[ 4 m_\ell^2 \ga \vel A \ver ^2 - \vel C
\ver ^2 \dr + m_B^2 s ( 1+ v^2 z^2) \ga \vel A \ver ^2 +\vel C \ver ^2 \dr
\Bigg]+ \nnb \\
&+& \frac{\lambda m_B^4}{2 r} \Bigg[ \lambda m_B^2 (1-v^2 z^2) \ga \vel B_2
\ver ^2 + \vel D_2 \ver ^2 \dr + 4 m_\ell^2 \vel D_2 \ver^2 (2+ 2 r -s ) \Bigg] + \nnb \\
&+& \frac{1}{2 r} \Bigg[ m_B^2 \left\{ \lambda (1- v^2 z^2) + 8 r s\right\}
\ga \vel B_1
\ver^2 + \vel D_1 \ver^2 \dr  
+ 16 m_\ell^2 r \ga \vel B_1 \ver ^2 - 2 \vel D_1 \ver ^2 \dr - \nnb \\
&-& 2 \lambda m_B^4
(1-r-s)(1-v^2 z^2)  \left\{ Re\ga  B_1 B_2^*  \dr + Re \ga D_1 D_2^*  \dr
\right\} - 16 \lambda m_B^2 m_\ell^2 
Re \ga D_1 D_2^* \dr \Bigg] + \nnb \\
&+&  \frac{2 \lambda m_B^2 m_\ell^2}{r} \Bigg[ s m_B^2 \vel D_3 \ver ^2 +
 2 Re \ga D_1 D_3^*  \dr - 2 (1-r) m_B^2 Re \ga D_2 D_3^*  \dr \Bigg] + \nnb \\
&+& \frac{2 \lambda m_B^6 s}{m_b^2} \left( \vel F_1 \ver ^2 v^2 + \vel F_2
\ver ^2 \right) + \nnb \\
&+&  \frac{4 \lambda m_B^3 m_\ell}{m_b \sqrt r} \Bigg[ Re\ga F_2 D_1^* \dr 
 - (1-r) m_B^2  Re  \ga F_2 D_2^* \dr
+ m_B^2 s  Re\ga F_2 D_3^* \dr 
 \Bigg] - \nnb \\
&-& m_B^2 \lambda^{1/2} z \Bigg[ 4 m_B \frac{m_\ell v}{m_b \sqrt r} \left\{ m_B^2
 Re \ga B_2 F_1^* \dr  - (1-r-s) Re\ga  B_1 F_1^* \dr \right\} + \nnb \\
&+& 8 m_B^2 s \left\{ v Re\ga  B_1 C^* \dr + Re\ga A D_1^*\dr \right\} \Bigg] \Bigg\}~,
\eea
where $z=cos \theta$\,, $\theta$ is the angle between the momentum of the $\ell$ lepton
and that of the $B$ meson in the center of mass frame of the lepton pair,
$\lambda = 1+r^2+s^2 -2 r - 2 s - 2 r s$, $r =
\frac{\ds{m_{K^*}^2}}{\ds{m_B^2}}$, 
$s=\frac{\ds{q^2}}{\ds{m_B^2}}$  and $v = \sqrt {1-\frac{\ds{4
m_\ell^2}}{\ds{q^2}}}$, $m_\ell$ are
the  lepton 
velocity and mass respectively.

If we put $F_1 = 0 $ and $F_2 = 0$ (which corresponds to the case where the 
contributions of the neutral Higgs boson exchange diagrams are neglected) we
get the results of \cite{R27}. As we noted earlier, the forward-backward
asymmetry and final lepton polarization asymmetry are very useful tools
for extracting more precise information on the Wilson coefficients
$C_7,~C_9^{eff}$ 
and $C_{10}$ and also for searching new physics.
Therefore in this work we
shall study these quantities as well. The forward-backward asymmetry $A_{FB}$
is defined in the following way:
$$ A_{FB} (q^2) = \frac{\displaystyle{\int_0^1 dz \frac{d \Gamma}{dq^2 dz} - \int_{-1}^0dz
\frac{d \Gamma}{dq^2 dz}}}{\displaystyle{\int_0^1 dz \frac{d \Gamma}{dq^2 dz}+
\int_{-1}^0dz\frac{d \Gamma}{dq^2 dz}}}~.$$

Let us now discuss the final lepton polarization. We define the
following three orthogonal unit vectors:
\bea
\vec{e}_L &=& \frac{\vec{p}_1}{\vel \vec{p}_1 \ver}~, \nnb \\
\vec{e}_T&=& \frac{\vec{p}_{K^*} \times \vec{p}_1}
{\vel \vec{p}_{K^*} \times \vec{p}_1 \ver}~, \nnb \\
\vec{e}_N &=& \vec{e}_T \times \vec{e}_L~, \nnb
\eea
where $\vec{p}_1$ and $\vec{p}_{K^*}$ are the three momenta of the
$\ell^-$
and the $K^*$ meson, respectively, in the center of mass of the
$\ell^+~\ell^-$ system. The differential decay rate for any given spin
direction $\vec{n}$ of the $\ell^-$ lepton, where $\vec{n}$ is a unit vector
in the $\ell^-$ lepton rest frame, can be written as
\bea
\frac{d \Gamma \ga \vec{n} \dr}{d q^2} = 
\frac{1}{2} \ga \frac{d \Gamma}{d q^2} \dr_{\!\!\! 0} 
\Big[ 1 + \ga P_L\, \vec{e}_L + P_N\, \vec{e}_N + P_T\, \vec{e}_T \dr \cdot
\vec{n} \Big]~,
\eea
where the subscript "0" corresponds to the unpolarized case and it can be
obtained from eq. (17) by integration over z. From definition
of $\vec{e}_i$ it is obvious that $P_T$ lies in the decay plane whose
orientation is organized by the vectors $\vec{p}_1$ and $\vec{p}_{K^*}$, and
$P_N$ is perpendicular to this plane.

The polarization components $P_i~(i=L,~T,~N)$ are defined as:
\bea
P_i (q^2) = \frac{ {\displaystyle{\frac{d \Gamma}{dq^2}
\ga \vec{n}=\vec{e}_i \dr -
\frac{d \Gamma}{dq^2}\ga \vec{n}=-\vec{e}_i \dr}} }
{ {\displaystyle{\frac{d \Gamma}{dq^2}\ga \vec{n}=\vec{e}_i \dr +
\frac{d \Gamma}{dq^2}\ga \vec{n}=-\vec{e}_i \dr}} } ~.
\eea

After standard calculations for $P_i \ga i=L,~T \dr$ we get 
\bea
P_L &=& \frac{v}{\Delta} \Bigg\{ \frac{32}{3}\lambda m_B^6  Re \ga A C^* \dr +
 \frac{4}{3} \frac{\lambda^2 m_B^6}{r} Re \ga B_2 D_2^* \dr +
4 \frac{m_B^2}{r}\left[ \frac{\lambda}{3} + 4 r s
\right] Re \ga B_1 D_1^* \dr - \nnb \\
&-& \frac{4}{3} \frac{\lambda m_B^4}{r} \ga 1 - r -s  \dr  
\left[ Re \ga B_2 D_1^* \dr + Re \ga B_1 D_2^* \dr \right] - 
8 \frac{\lambda m_B^6 s}{m_b^2} Re \ga F_1 F_2^* \dr -\nnb \\
&-& 8 \frac{\lambda m_B^3 m_\ell}{m_b \sqrt{r} }
\left[ Re \ga F_1 D_1^* \dr -  \ga 1-r \dr Re \ga F_1 D_2^* \dr +
 m_B^2 s Re \ga F_1 D_3^* \dr \right] \Bigg\}~,  \\ \nnb \\
P_T &=& \frac{\sqrt{\lambda} \pi}{\Delta} \Bigg\{
\frac{\lambda m_B^5 m_\ell \ga 1-r \dr}{r \sqrt{s} } Re \ga B_2 D_2^* \dr
 + 8 m_B^3 m_\ell \sqrt{s} Re \ga A B_1^* \dr
- \lambda m_B^5 m_\ell \frac{\sqrt{s}}{r} Re \ga B_2 D_3^* \dr - \nnb \\
&-& \lambda m_B^6 \frac{\sqrt{s}}{m_b \sqrt{r}} Re \ga B_2 F_2^* \dr + 
m_B^3 m_\ell  \ga 1 -r-s \dr \frac{\sqrt{s}}{r} Re \ga B_1 D_3^* \dr + \nnb\\
&+& m_B^4 \ga 1-r-s \dr \frac{\sqrt{s}}{m_b \sqrt{r}} Re \ga B_1 F_2^* \dr -
m_B^3 m_\ell \ga 1-r-s \dr \frac{\ga 1-r\dr}{r \sqrt{s}} Re \ga B_1 D_2^* \dr
+ \nnb \\
&+& m_B m_\ell \frac{ \ga 1-r-s \dr }{r \sqrt{s}} Re \ga B_1 D_1^* \dr -
\lambda m_B^6 \sqrt{s} \frac{v^2}{m_b \sqrt{r}} Re \ga D_2 F_1^* \dr + \nnb \\
&+& m_B^4 \sqrt{s} \ga 1-r-s \dr \frac{v^2}{m_b \sqrt{r}} Re \ga D_1 F_1^* \dr
- \lambda m_B^3 \frac{m_\ell}{r \sqrt{s}} Re \ga B_2 D_1^* \dr  
\Bigg\}~.
\eea
The denominator $\Delta$ in eqs.(20) and  (21) can be obtained from eq.(17) by
integration over $z$ of the terms within the curly bracket. 
We also calculate the $P_N$ component of the  lepton polarization, but
the numerical results show that its value is  quite small and because of
that we do not present its explicit form here.

\section{Numerical Analysis}
The values of the main input parameters, which appear in the expression for the decay width
are: $m_b=4.8~GeV,~m_c=1.35~GeV,~m_\tau = 1.78~GeV,~m_\mu = 0.105~GeV,
~\Lambda_{QCD} = 225~ MeV,~m_B = 5.28~GeV$,
and $m_{K^*}=0.892~GeV$. We use the pole form of the formfactors given in eq.(16). For $B$ meson
lifetime we take $\tau(B_d)=1.56 \times 10^{-12}~s$ \cite{R33}. The values of the Wilson
coefficients $C_7^{SM} (m_b)$ and $C_{10}^{SM} (m_b)$ to the leading logarithmic
approximation are \cite{R34,R35}:
$$ C_7 = -0.315~,~~~C_{10} = -4.642~.$$
The expression $C_9^{eff}$ for the $b \rar s$ transition in the next to
leading order approximation is given as (see for example \cite{R34})
\\ 
\bea
\lefteqn{
C_9^{eff} (m_b) = } \nnb \\
&& C_9^{SM} (m_b) + C_9^{H^-} (m_b) + 0.124 w( \hat{s}) + g( \hat{m_c},\hat{s}) \left( 3 C_1 +
C_2 + 3 C_3 + C_4 + 3 C_5 + C_6 \right) - \nnb \\
&-&\frac{1}{2} g( \hat{m_q},\hat{s})
\left( C_3 + 3 C_4 \right) - \frac{1}{2} g( \hat{m_b},\hat{s}) \left( 4C_3 +
4 C_4 + 3 C_5 + C_6 \right) + \nnb \\
&+& \frac{2}{9} \left( 3 C_3 + C_4 + 3 C_5 + C_6
\right)~,
\eea  
with $$ C_1 = -0.249~,~~ C_2 = 1.108~,~~C_3 = 1.112 \times 10^{-2}~,~~ C_4
= -2.569 \times 10^{-2}~,$$ $$ C_5 = 7.4 \times 10^{-3}~,~~C_6 = -3.144
\times 10^{-2}~,~~C_9^{SM} (m_b) = 4.227,$$ where $\hat{m_q}=\frac{\ds{m_q}}{\ds{m_b}},~ \hat{s} =
\frac{\ds{q^2}}{\ds{m_b^2}}$.

In the above expression, $ w(\hat{s})$ represents the one gluon correction to
the matrix element $O_9$  and its explicit
form  can be found in \cite{R13},
while the function $g( \hat{m_q},\hat{s})$ arises from the one loop
contributions of the four quark operators $O_1 - O_6$ (see for example
\cite{R34,R35}), i.e.
\bea
g(\hat{m_q},\hat{s^\prime}) &=& - \frac{8}{9} ln\, \hat{m_q} + \frac{8}{27} +
\frac{4}{9} y_q - \frac{2}{9} ( 2 + y_q) \sqrt { 11 - y_q} + \nnb \\
&+& \left\{ \theta (1-y_q) \left( ln \frac{1+ \sqrt{1-y_q}}{1-
\sqrt{1-y_q}} - i \pi \right) + \theta(y_q -1) arctan \frac{1}{\sqrt{y_q -1}}
\right\}~, 
\eea
where $y_q = \frac{\ds{\hat{{m_q}}}}{\ds{\hat{s^\prime}}}$, and $\hat{s^\prime} =
\frac{\ds{4
q^2}}{\ds{m_b^2}}$

In all numerical calculations we use the following values for the
masses of the Higgs particles: $m_{H^\pm} =
200~GeV,~m_{h^0}=80~GeV,~m_{H^0}=150~GeV$, and $m_{A^0} = 100~GeV$. 
We also take $sin \alpha = \frac{\ds{\sqrt 2}}{\ds{2}}$ and for
$tan\,\beta$ we choose the following set of values: 
$tan \, \beta = 1,~tan\, \beta = 30$ and $tan \, \beta =50$.

In Fig.1 we present the $q^2$ dependence of the differential branching ratio
for $B_d \rar K^* \tau^+ \tau^-$ decay with and without the long distance
effects. It follows from this figure that the differential branching ratio
is sensitive to the value of $tan \,\beta$ if the $\psi^\prime$ mass region is
excluded. For example at $tan \, \beta=50$ the differential branching ratio is
approximately 2 times larger than the one at $tan \, \beta=1$.

In Fig.2 we plot the dependence of the forward-backward asymmetry $A_{FB}$ 
on $q^2$ with and without the long-distance effects at different values of
$tan \, \beta$. From this figure we see that $A_{FB}$ is positive for all
values of $q^2$ except in the $\psi^\prime$ resonance region and it is sensitive
to the value of $tan \, \beta$.
 
In Fig.3 we depicted the $q^2$ dependence of the longitudinal polarization
of the final lepton $P_L$ with and without the long distance effects at
different values of $tan \, \beta$. As we see from this figure if we exclude the
resonance mass region of $\psi^\prime$, $P_L$ is negative for all values of
$q^2$ at $tan \, \beta=1$ as well as at  $tan \, \beta=30$ and $tan \,
\beta=50$.

In Fig.4 we present the $q^2$  dependence of the transversal polarization
$P_T$ of the $\tau$ lepton which lies in the decay plane, without long
distance effects at $tan\, \beta=1$, $tan\, \beta=30$ and at  $tan \, \beta=50$. From this
figure it follows that at $tan\, \beta=1$  $P_T$ 
is positive at all values of $q^2$. For $tan \, \beta=30$ and $tan
\,\beta=50$, $P_T$ is positive near the threshold region but far from the
threshold it becomes negative. 
Therefore the determination of the sign of $P_T$ in future experiments is a
very important issue and can provide a direct information   
for the establishment of new physics.

In Fig.5 we present the differential branching ratio versus $q^2$ for \tepm.
From this figure and Fig.1 it follows that the differential branching ratio for
the \tepm decay is approximately one order of magnitude greater then that of the \tept
decay. Note that for $tan \, \beta = 30$ and $tan \, \beta = 50$, the
results for the differential branching ratio are practically the same.

In Fig.6, we depict the dependence of $A_{FB}( B \rar K^* \mu^+ \mu^-)$ on
$q^2$ at $tan \, \beta = 1$ and $tan \, \beta = 50$. From a comparison  of
this figure with Fig.2 we observe that the behaviour of $A_{FB}$ for the \tepm
decay is totaly different from that for the \tept decay.
In the \tept case $A_{FB}$ is positive for all values
of $q^2$ (without the long distance effects), while it changes its sign here in the \tepm case.
 This indicates that for the \tept
case, the neutral Higgs boson exchange diagrams give considerable
contribution. It follows from these 
two figures then that the determination of the sign of $A_{FB}$ in
different kinematical regions is also a very important tool for establishing
new physics.

The behaviour of the longitudinal polarization with changing $q^2$ is presented in Fig.7.
In \tepm decay, $P_L$ is firstly positive and then it changes sign 
(without the long distance effects), which is
absolutely different from that of the \tept case, for which $P_L$ is negative for
all $q^2$.

Since, as we have already noted, $A_{FB},~P_L$, and $P_T$ contain
independent information, their investigation in experiments will be a very
efficient tool in establishing new physics.
 
At the end of this section we present the values of the branching ratios
for the $B_d \rar K^* \tau^+ \tau^-$ decay.
After integrating over $q^2$ we get for  the branching ratios for the 
$B_d \rar K^* \tau^+ \tau^-$ decay, without the long distance contributions,
\bea
B(B_d \rar K^* \tau^+ \tau^-) = \left\{ \begin{array}{ll}
~ 0.86 \times 10^{-7}   & (\tan \, \beta = 1) \\ \\
~ 0.90 \times 10^{-7}   & (\tan \, \beta = 30) \\ \\
~ 1.11 \times 10^{-7}   & (\tan \, \beta = 50)~.
\end{array} \right.
\eea

The ratio of the exclusive and inclusive channels is defined as
$$R = \frac{\displaystyle{B(B_d\rar K^* \tau^+ \tau^-)}}
         {\displaystyle{B(b \rar s \tau^+ \tau^-)}}~.$$
In the SM this ratio is given as $R = 0.270 \pm 0.007$ when
$B(b \rar s \tau^+ \tau^-) = (2.6 \pm 0.5) \times 10^{-7}$ \cite{R35}.
In our case we have 
\bea
R  = \left\{ \begin{array}{ll}
~ 0.33   & (\tan \ \beta = 1) \\ \\
~ 0.35   & (\tan \,\beta = 30) \\ \\
~ 0.43   & (\tan \, \beta = 50)~,
\end{array} \right.
\eea
where we use the SM value for the inclusive $B(b \rar s \tau^+ \tau^-)$.

In conclusion, we calculate the rare $B \rar K^* \ell^+ \ell^-$ decay in 
2HDM. It is observed that $A_{FB}$, $P_L$ and the transversal 
polarization $P_T$ of the charged lepton are very sensitive to the value 
of $tan \, \beta$. Therefore, in search of new physics, their experimental 
investigation can serve as a crucial test. 

\newpage

\section*{Figure Captions}
In Fig.1, Fig.2 and Fig.3,  lines 1, 3 and 4
correspond to the short distance contributions for the values of $tan \, \beta
= 50$, $tan \, \beta= 30$ and $tan \, \beta= 1$, respectively; while lines
2, 5
and 6 correspond to the sum of the short and long distance contributions for
the values of $tan \, \beta= 50$, $tan \, \beta= 30$ and $tan \, \beta=
1$, respectively. \\ \\
{\bf 1.} 
Invariant mass squared ($q^2$) distribution of the $\tau$ lepton       
pair for the decay $B \rar K^* \tau^+ \tau^-$.\\ \\
{\bf 2.} The dependence of the forward-backward asymmetry $A_{FB}$ on $q^2$
 for the decay $B \rar \\ \indent  K^* \tau^+ \tau^-$. \\ \\
{\bf 3.} The dependence of the longitudinal polarization $P_L$ on
$q^2$ for the $B \rar K^* \tau^+ \tau^-$. \\ \\
{\bf 4.} The same as in Fig.3 but for the transversal 
polarization $P_T$. \\ \\
{\bf 5.} The same as in Fig.1, but for the 
         $B \rar K^* \mu^+ \mu^-$ decay. \\ \\
In Fig.6 and Fig.7, lines 1 and 3 correspond to the short distance
contributions for the values of $tan \, \beta= 50$ and $tan \,
\beta=1$, respectively; while lines 2 and 4 correspond to the sum of the
short and long distance contributions for the values of $tan \, \beta= 50$
and $tan \, \beta=1$, respectively. \\ \\
{\bf 6.} The same as in Fig.2, but for the 
         $B \rar K^* \mu^+ \mu^-$ decay. \\ \\
{\bf 7.} The same as in Fig.3, but for the
         $B \rar K^* \mu^+ \mu^-$ decay.

\begin{figure}
\vspace{25.0cm}
    \includegraphics{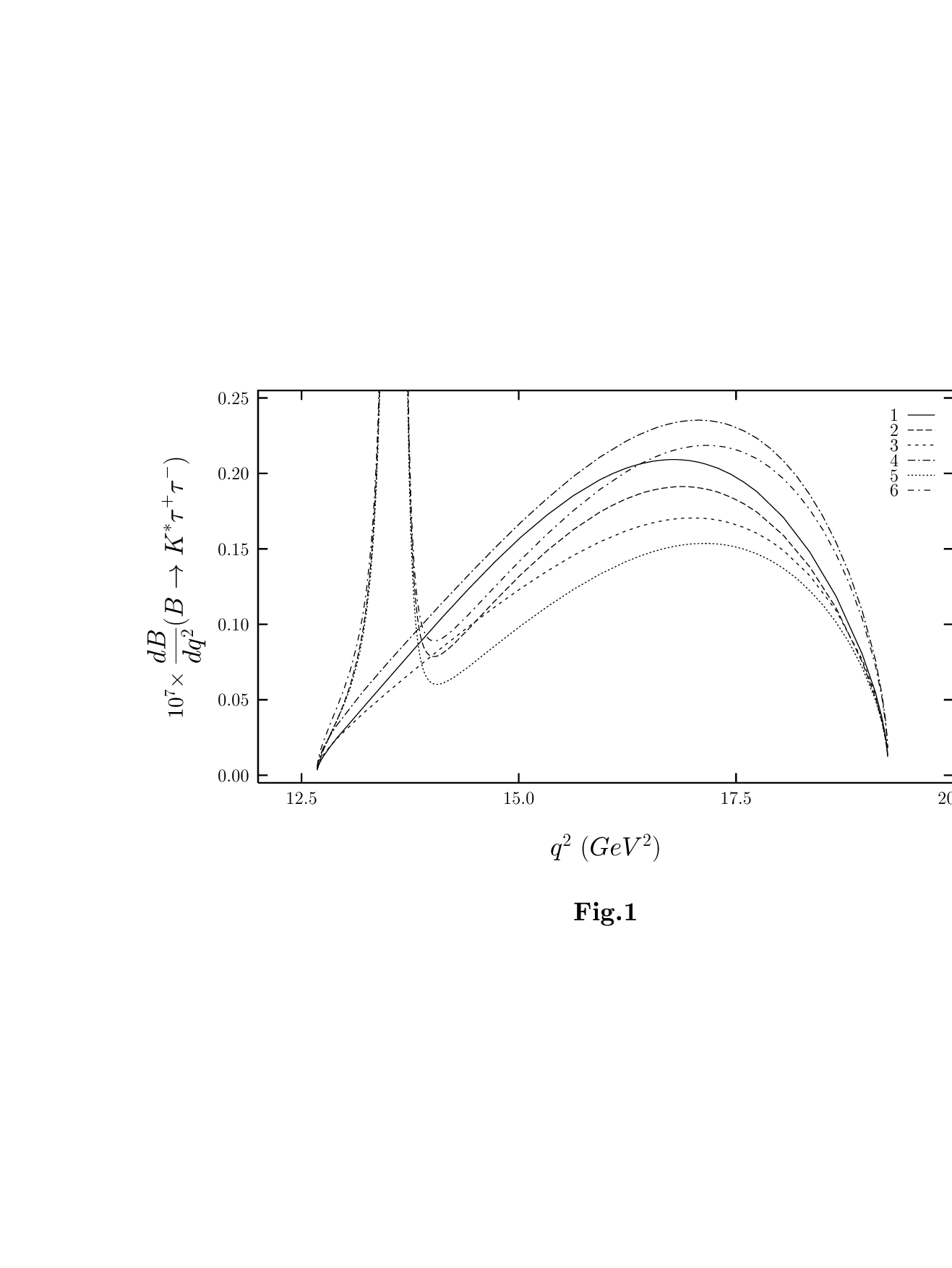}
    \includegraphics{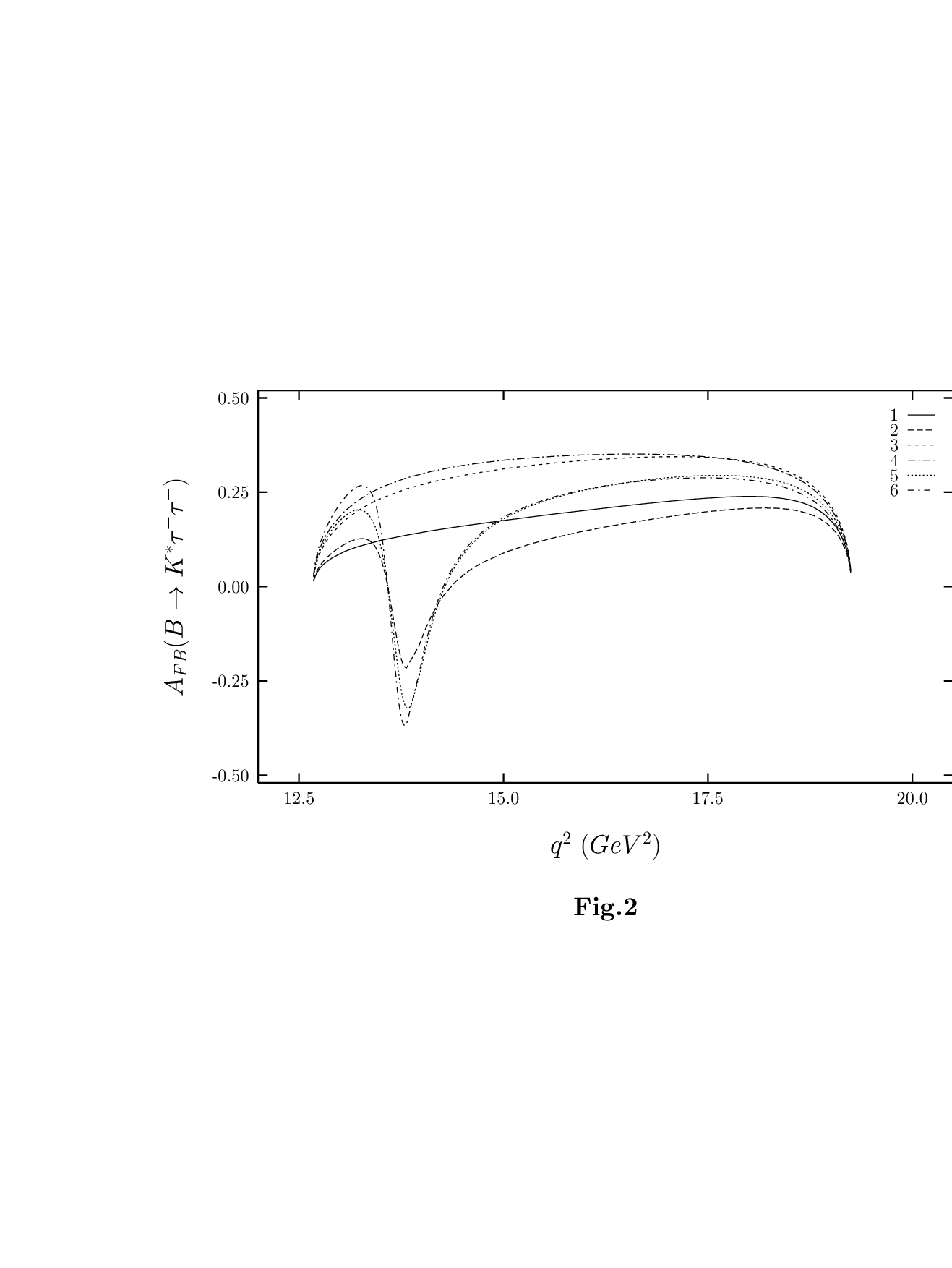}
    \vspace{-4.0cm}
\vspace{0.0cm}
\end{figure}

\begin{figure}
\vspace{25.0cm}
    \includegraphics{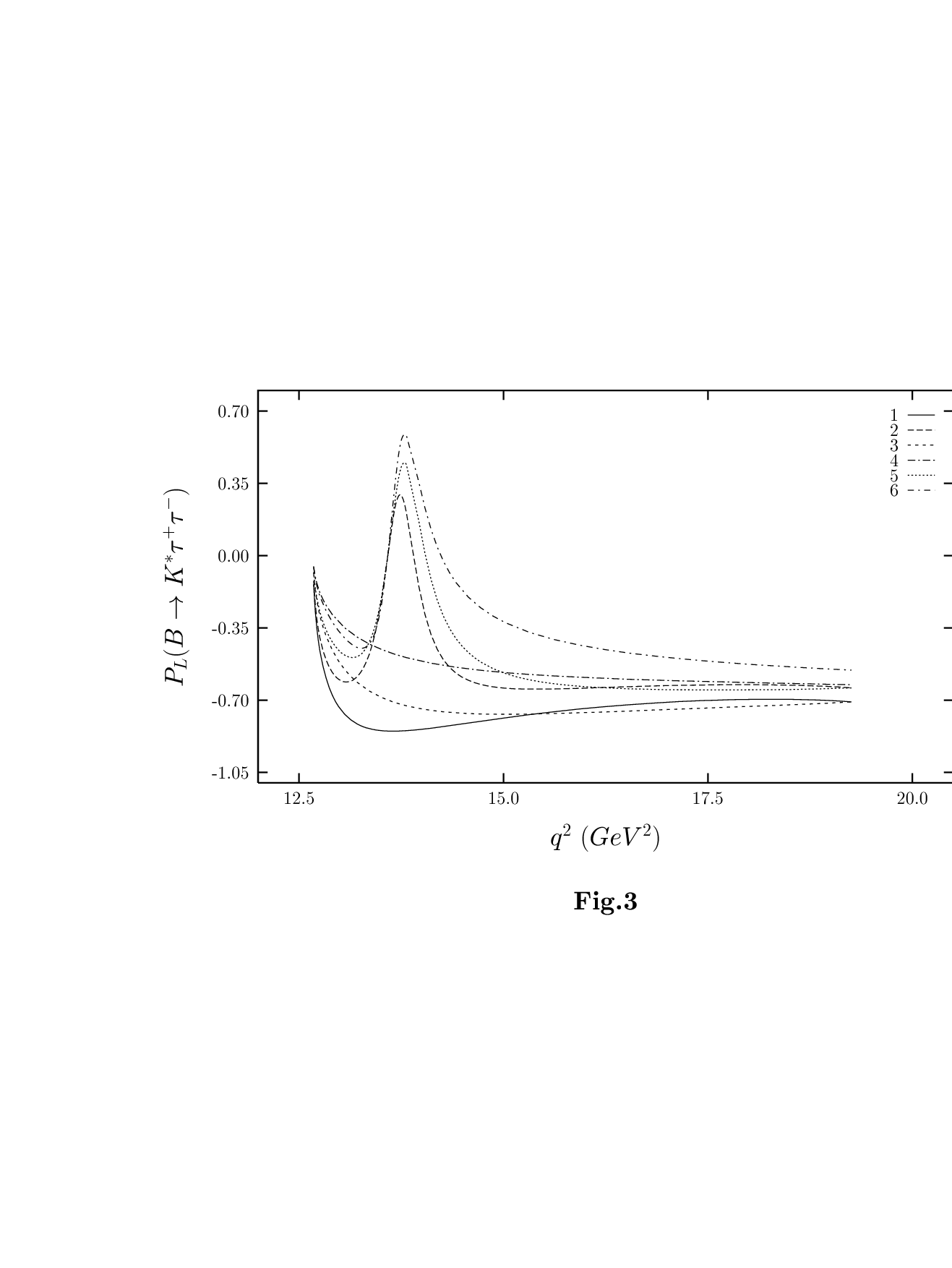}
    \includegraphics{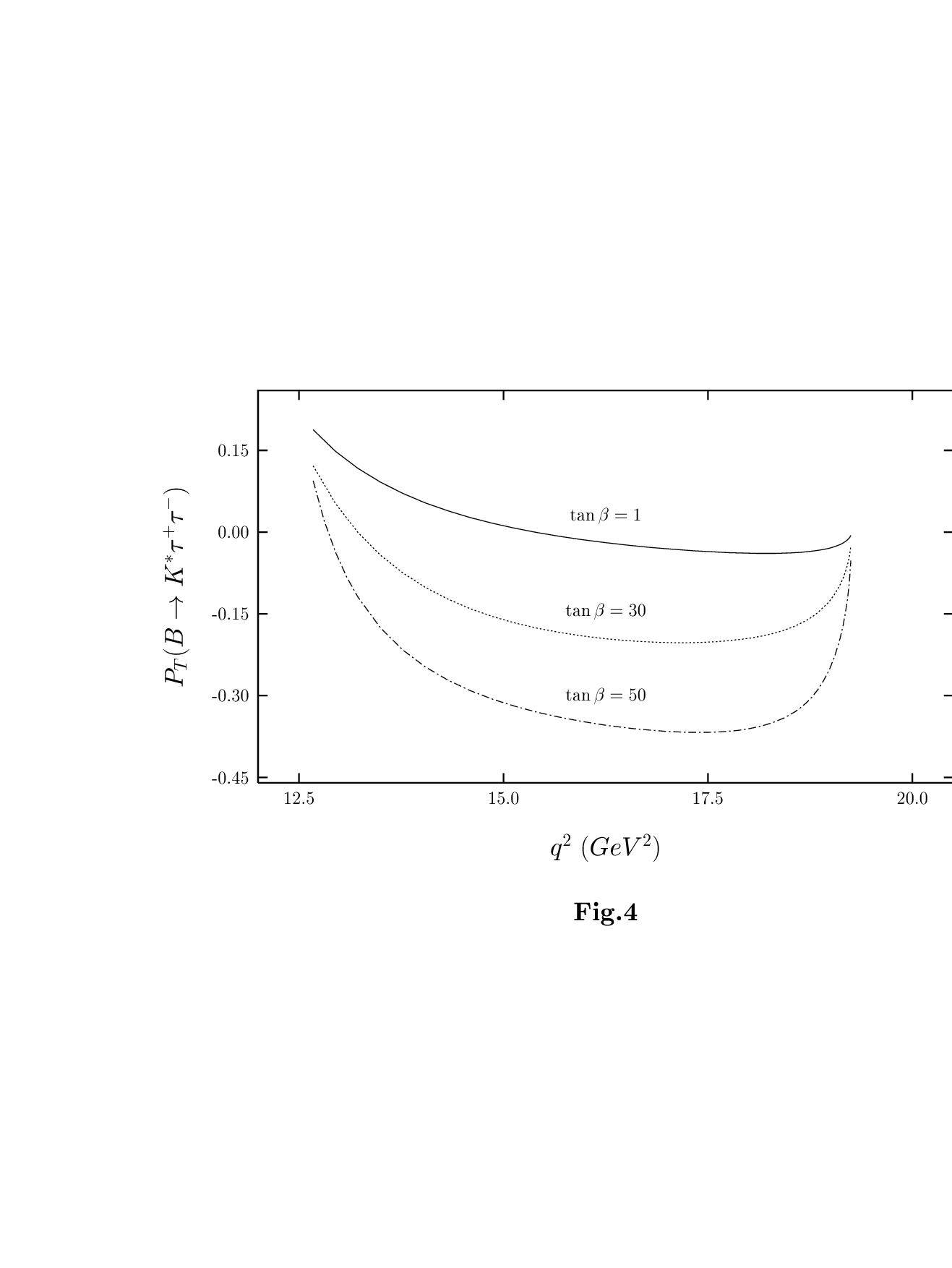}
    \vspace{-4.0cm}
\vspace{0.0cm}
\end{figure}

\begin{figure}
\vspace{25.0cm}
    \includegraphics{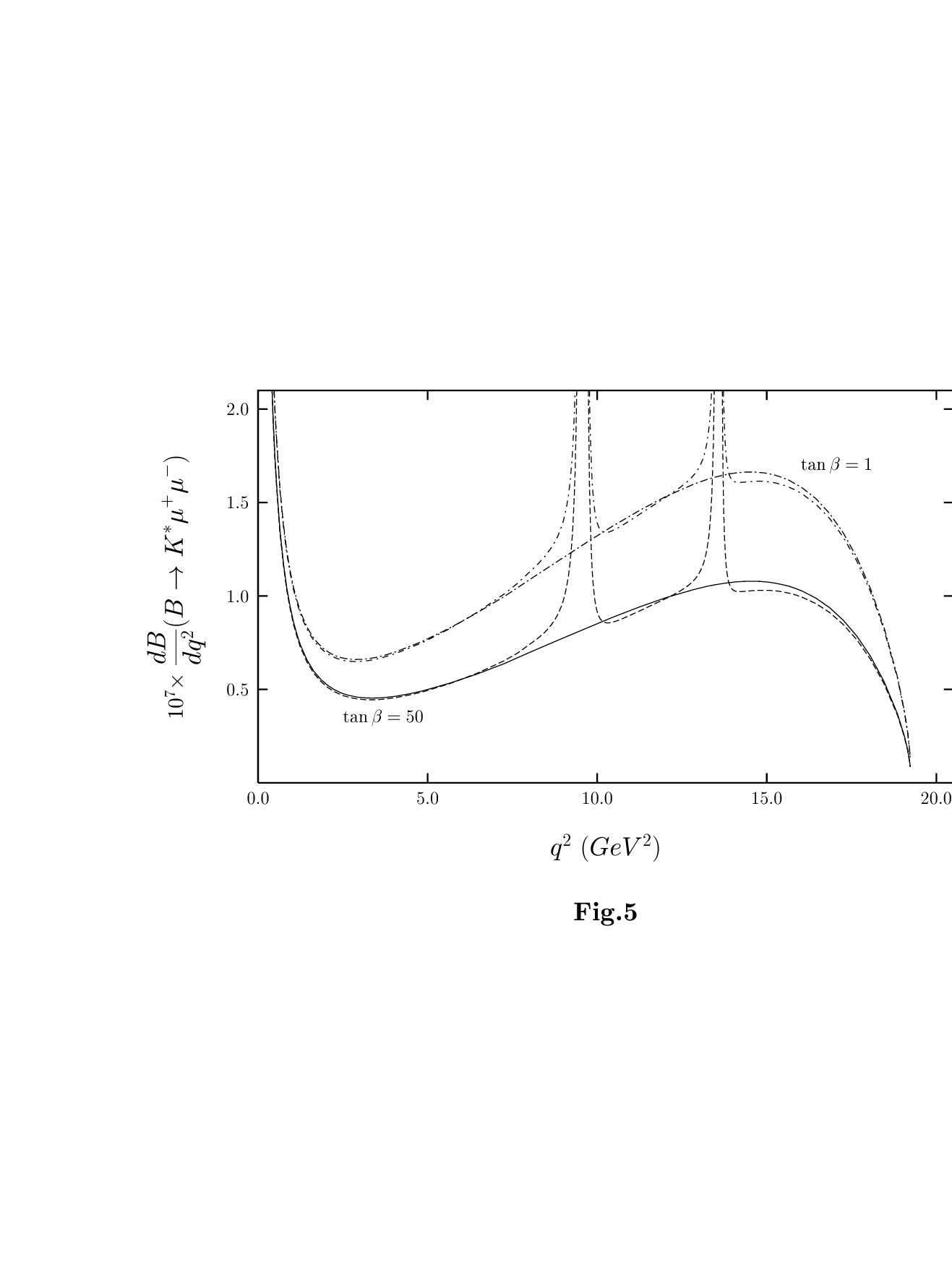}
    \includegraphics{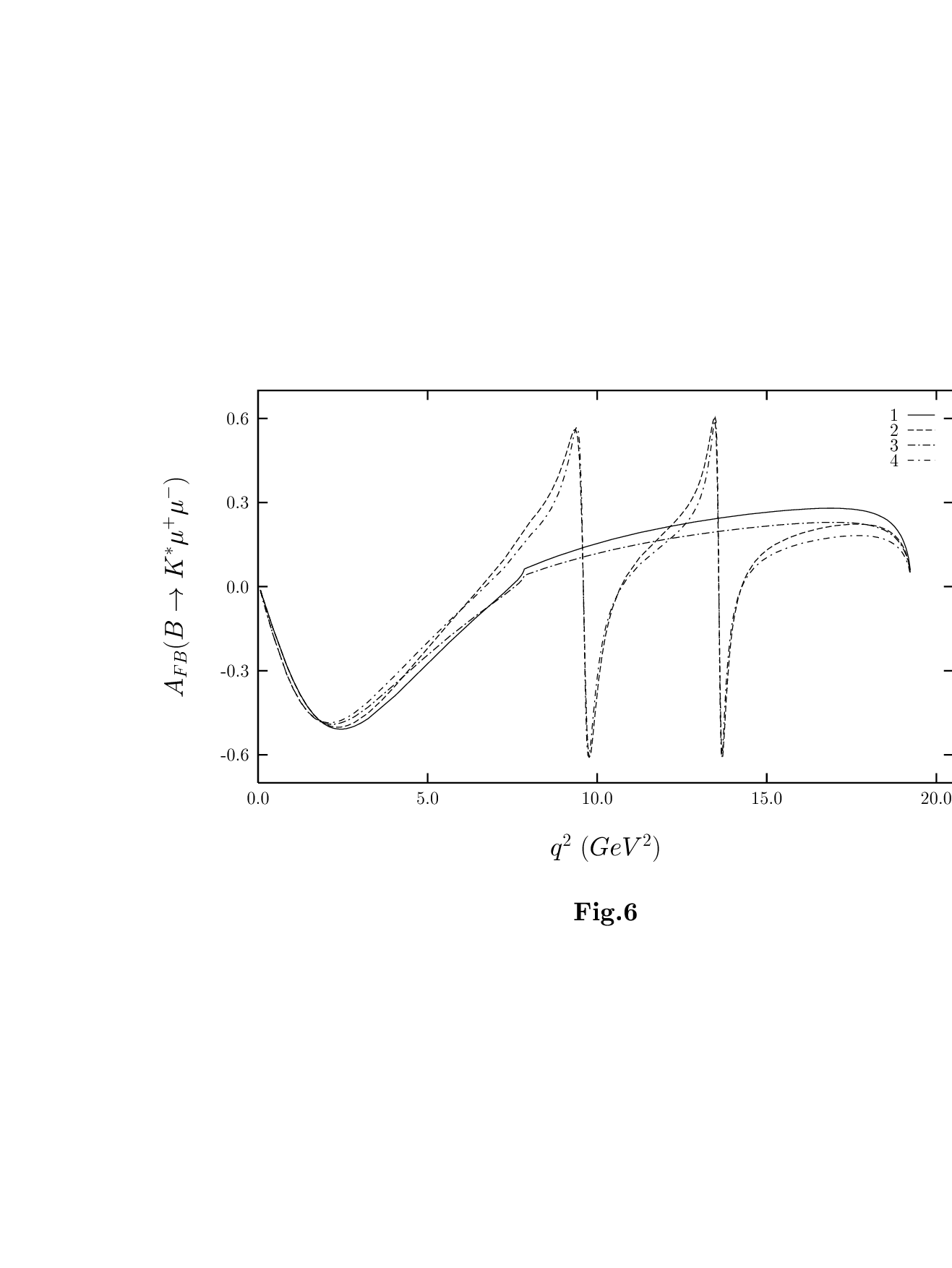}
    \vspace{-4.0cm}
\vspace{0.0cm}
\end{figure}  

\begin{figure}
\vspace{25.0cm}
    \includegraphics{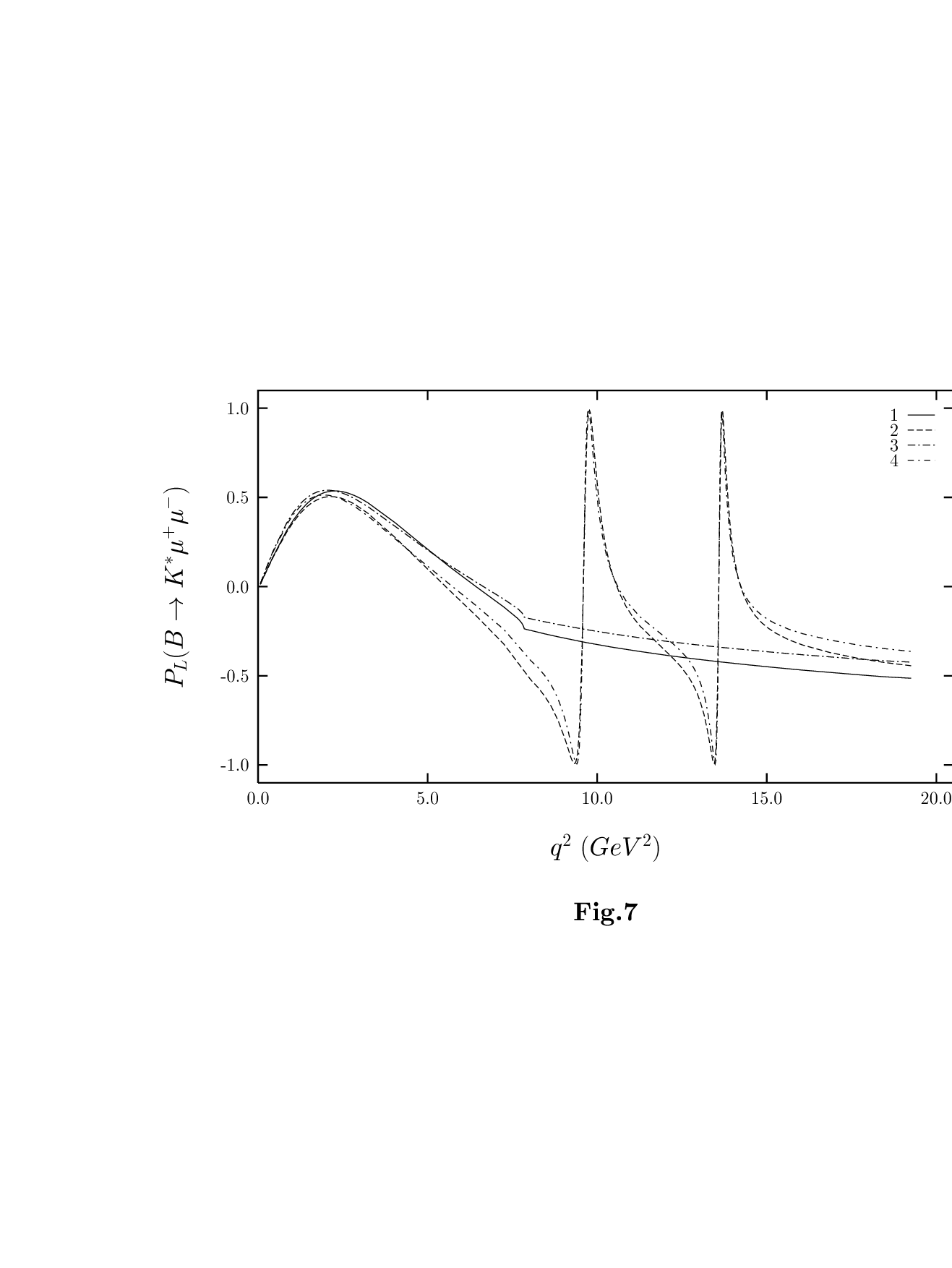}
    \vspace{-4.0cm}
\vspace{0.0cm}
\end{figure}

\end{document}